# Student perception and the efficacy of universities in shaping the entrepreneurial mindset

ANDREA S. GUBIK, PHD
ASSOCIATE PROFESSOR

UNIVERSITY OF MISKOLC
e-mail: getgubik@uni-miskolc.hu

ZOLTÁN BARTHA, PHD
ASSOCIATE PROFESSOR

UNIVERSITY OF MISKOLC
e-mail: zoltan.bartha@ekon.me

*SUMMARY*

*Modern universities may play a significant role in entrepreneurial ecosystems by boosting the entrepreneurial activity of the region. One way to achieve this is through entrepreneurship education. In this study we suggest that one reason why entrepreneurship education has a weak impact on entrepreneurial activity is that the effect of courses and extracurricular programmes depends on how students perceive the entrepreneurial activity. We use the 2018 GUESSS database, which includes 9,667 answers for Hungary, to develop a general linear model. The model suggests that students' entrepreneurial intentions, attitudes toward entrepreneurship, self-efficacy, social norms, as well as the university, and the field of study all have a small but statistically significant impact on how students perceive the entrepreneurial ecosystem within the university. Our conclusion is that more emphasis on shaping attitudes and arousing student interest can increase the efficiency of entrepreneurship education.*
*Journal of Economic Literature (JEL) codes: L26, I23*
*DOI: http://doi.org/10.18096/TMP.2021.01.07*

## INTRODUCTION

Although entrepreneurs have attracted considerable research interest for at least half a century, entrepreneurship moved into the core of business research in the 21st century. Koppl et al. (2015), when attempt to create a framework for a non-stable and non-causal economic system, coin the notion of entrepreneurial economics. In their approach "entrepreneurs solve the frame problem of social systems", where the frame problem can be defined as "the problem of modifying the system's implicit frame of analysis to adapt successfully to non-algorithmic change" (p. 22). In other words, entrepreneurs are the ones who make it possible for the economic system to adapt to unforeseeable changes, and so they have a fundamental impact on its performance and its ability to create and provide wealth to the members of the community.

Baumol et al. (2007) suggest that developing entrepreneurial skills is increasingly important for growth and prosperity; one of the current arguments is that the main reason for this lies in machine learning and the rise of artificial intelligence (BrynjoIfsson & McAfee 2014; Acemoglu & Restrepo 2018). The rise in the demand for entrepreneurial skills is supported by empirical evidence as well (e.g. Prüfer & Prüfer 2020). One of the key takeaways of the empirical studies is that entrepreneurial skills are not only vital for classical entrepreneurs; they represent a meta-capability that is also required in large organisations (intrapreneurship), or in areas where social change should be promoted (social entrepreneurship) (Obschonka et al. 2017).

Entrepreneurial thinking and entrepreneurial skills are thus seen as 21st century skills. Another line of literature focused on entrepreneurial ecosystems (Moore 1993; Zacharakis et al. 2003, Mason & Brown 2013) suggests that universities have a role in training entrepreneurial skills. The research problem addressed in this paper is related to this: we examine how effective universities can be at training students with little interest in entrepreneurship, and entrepreneurial skills.

We use the 2018 GUESSS (Global University Entrepreneurial Spirit Students' Survey) database to test our research question. The novelty of our discussion lies in the incorporation of student perception in our analysis: we may get a more accurate picture on the effectiveness of university projects aimed at training entrepreneurial skills, when the attitude of students towards entrepreneurship is also considered.

# LITERATURE REVIEW

We need to review three areas in order to formulate a testable hypothesis on entrepreneurship education: the literature of entrepreneurial ecosystems that connect universities with entrepreneurship; theories and empirical tests on what impact universities have on ecosystems; and finally, the literature of perception as an influence in training performance.

In his 1993 article Moore suggested that "a company be viewed not as a member of a single industry but as part of a business ecosystem that crosses a variety of industries" (Moore 1993, p. 76). When new innovations are born, companies not only compete, they also work cooperatively, and the business ecosystem is the framework through which this coevolution takes place. Business ecosystems are typically defined as a set of large number of hubs (entrepreneurial actors, organisations and processes, and different institutions) that are loosely interconnected in a complex way (Mason & Brown 2013; Ma 2019). Over the past decade (2012 is the first year when the number of academic papers related to the topic exceeds 10 in the Scopus database; see Figure 1 in Cavallo et al. 2019, p. 1293) a number of sophisticated business (or 'entrepreneurial', as they are more commonly called nowadays) ecosystem models have been developed, and universities typically are included in them. Isenberg (2011) created a complex model where six major domains interact; universities are part of it through the Human Capital domain. A later model published by the World Economic Forum (WEF 2014), created by a team that included Isenberg, Foster, and Shimizu, listed eight pillars of an entrepreneurial ecosystem; one of these pillars is formed by major universities as catalysts of the entrepreneurial process. The 6+6 model suggested by Koltai & Muspratt (2016) has six pillars and six key actors, that contribute to the development and training of entrepreneurial skills, one of which is academia.

Even though universities are considered natural actors in entrepreneurial ecosystems, their exact role and impact on the system is disputed. One obvious role is to provide talent for the entrepreneurial process (e.g. Isenberg 2011; Feld 2012). The World Economic Forum (2014) suggests that universities have an even deeper role in promoting the entrepreneurial culture within the ecosystem, and they also play a key role in idea-formation. This latter is questioned by Feld (2012) when criticising the intellectual property rights regulations that universities impose on their spin-off firms. Åsterbro & Bazzazian (2011) add that the number of spin-offs created by the median university among top US universities is quite low (less than two per year).

Koltai & Muspratt (2016) believe that universities play a key role in promoting entrepreneurial skills. Isenberg (2014), on the other hand, claims that entrepreneurship education might be helpful, but it is not critical for the success of the entrepreneurial ecosystem (arguing that the first master course in technological entrepreneurship in Israel started in 1987, 15 years after the first Israeli initial public offering at NASDAQ). Rice et al. (2014) provide very detailed case studies on six successful university-based entrepreneurial ecosystems. They find that strong faculty leadership and constant curriculum development are both key success factors.

Given the critical role of universities in (at least some) entrepreneurial ecosystems, a new line of studies have emerged related to the so-called entrepreneurial universities (Guerrero et al., 2006, 2015; Sánchez-Barrioluengo et al., 2019). Entrepreneurial universities contribute to social and economic development not only through their traditional teaching and research activities, but through a third pillar as well, connected to the boosting of entrepreneurial activities. In this paper we focus on one element of this entrepreneurial pillar: the promotion of entrepreneurial skills and values. This may be achieved through curriculum development and through other activities that influence student perceptions on entrepreneurship.

Entrepreneurship education is a very fast-growing field within the higher education domain. In 1985 around 250 courses focused on entrepreneurship at colleges in the USA; by 2008 this number had grown to over 5,000 (Hayes et al., 2020). The impact of such courses is disputed. Isenberg (2014) does not see these courses as critical. Studies tend to confirm that university graduates have a higher intention of becoming entrepreneurs (e.g. İlhan Ertuna & Gurel 2011). Entrepreneurship education may enhance entrepreneurial intentions (Block et al. 2013; Barba-Sánchez & Atienza-Sahuquillo 2018), but it was also found that after completing an entrepreneurship course the entrepreneurial intention declines (von Graevenitz et al. 2010). The impact may be dependent on the content of the course (Bădulescu et al., 2020). In addition to the content, it is also important to pay attention to what methods and learning forms come to the fore: traditional frontal solutions that can be used in case of large numbers of students, or teaching taking place in small groups using interaction, cooperative techniques, simulation or case studies. Numerous studies argue for the effectiveness of the latter and emphasise that there is a need to move toward more unconventional, experienced-based teaching, and evaluation methods (Solomon et al. 1994; Kickul & Fayolle 2007; Harms, 2015; Costin et al. 2018). They complement and reinforce prior classroom learning through application, facilitate learning about the real world of the entrepreneur, and have a positive impact on entrepreneurial intentions (Mason & Arshed 2013) and on self-efficacy.

The efforts of universities are just one side of the coin. As the interpretation of the environment is based on prior knowledge about it, which consists partly of the individual's personal experience and partly of experience and knowledge taken from others (Farkas 2010), given the availability of the same conditions and information, there are significant differences in the perception of the usefulness of a program and in general there are differences in the recognition of opportunities among students. From the point of view of entrepreneurship, the role of an entrepreneurial family background seems to be important; according to the literature, this has an incentive effect on entrepreneurial intention (Gubik & Farkas 2019), as well as perceived social norms, which mediate possible roles and values. A supportive environment increases the entrepreneurial intention (Liñán & Chen 2009, Autio & Wennberg 2010). In a study of German university students, Bergmann et al. (2018) found that the students' perception of the entrepreneurial climate around the university is shared among students, and so they suggest that the entrepreneurial climate perceptions differ between students with and without an affinity for entrepreneurship.

All these factors (knowledge, experience, values) largely determine how students relate to entrepreneurship and how they evaluate their own role and skills. This affects not only how motivated they are to participate and how well they perform, but also how much they perceive the availability of opportunities. Perception is the result of a brain process based on the sensation of current events, which also uses our knowledge and previous experiences (Csépe 2017). We are only able to perceive a fraction of the continuous stimuli we experience (this selection process is attention) (Juhász & Takács 2007). Attention only processes those moments that are important for current behaviour (Schutz 1962; Csépe 2017) and is influenced by internal factors such as attitudes, expectations, values and beliefs. The consequence of this is relevant to this article: the supply side (i.e., the provision of services by the university) is not sufficient to achieve the objectives set. Gubik (2013) states that the task is not merely to increase the number of courses or to increase the provision and availability of other services and resources, but to get students to recognise their potential.

In the following, we examine whether there are differences in student assessment of the entrepreneurial ecosystem among universities and what factors cause the differences in evaluation within the same university. The former is attributed to the different university efforts and the different characteristics of the universities, while the latter is attributed to the different attitudes and interests of the students.

# DATA AND METHODS

GUESSS (Global University Entrepreneurial Spirit Students' Survey) investigates entrepreneurial intentions and activities of university students. The survey explores the students' career intentions, the families' and students' own businesses, and investigates their motivations and goals and their orientation and behaviour in their business activity. It also analyses the role of higher education and culture in the decision.

The first survey was conducted in 2003 with the participation of two countries. By 2018 55 countries had joined the project and 208,636 students sent their responses to the questionnaire. In the framework of this paper we investigate the sample of Hungary, where the 2018 database contains 9667 answers. Table 1 shows the distribution of the sample according to the main descriptive statistics of the respondents.

*Table 1*
*Distribution of GUESSS 2018 Survey Participants, Hungary*

| Study level | % | Gender | % | Area of studies | % |
|---|---|---|---|---|---|
| Bachelor | 69.6 | Female | 58.3 | Arts/Humanities (e.g., cultural studies, history, linguistics, philosophy, religion) | 8.4 |
| Master | 16.3 | Male | 41.7 | Business/Management | 10.6 |
| PhD | 2.4 | | | Computer sciences/IT | 9.3 |
| Other (e.g., MBA) | 11.7 | | | Economics | 11.7 |
| | | | | Engineering (incl. architecture) | 22.3 |
| | | | | Human medicine/health sciences | 8.3 |
| | | | | Law | 6.2 |
| | | | | Mathematics | 0.7 |
| | | | | Natural sciences | 6.5 |
| | | | | Science of art (e.g., art, design, dramatics, music) | 0.7 |
| | | | | Social sciences (e.g., psychology, politics, education) | 8.3 |
| | | | | Other | 7.1 |

Source: own elaboration, N=9667

About 22.3% of respondents studied engineering, 22.2% of students studied business and management and economics. Computer sciences/IT students represented 9.3% and Arts/Humanities students accounted for 8.4%. The sample of students in medicine, health sciences and social sciences amounted to 8.3%. The vast majority of respondents (70%) attended bachelor-level studies. The proportion of master-level students in the sample was much lower (16%). Regarding the respondents' gender, the sample contained a larger ratio of females (58.2%). As for age, 85.2% of respondents were born after 1990, that is, they were younger than 28 at the time of completing the questionnaire.

## Variables

On the basis of the survey of 2018, the entrepreneurial ecosystem and its influencing factors were analysed. We used the following variables:

*Entrepreneurial ecosystem (ECO)*

-Please indicate the extent to which you agree with the following statements about the university environment: The atmosphere at my university inspires me to develop ideas for new businesses; There is a favourable climate for becoming an entrepreneur at my university; At my university, students are encouraged to engage in entrepreneurial activities. (1-7 Likert scale)

*Entrepreneurial intention (INT)*

-Please indicate your level of agreement with the following statements: I am ready to do anything to be an entrepreneur! My professional goal is to become an entrepreneur; I will make every effort to start and run my own business; I am determined to create a business in the future; I have a very seriously thought of starting a business; I have a strong intention to start a business someday. (1-7 Likert scale)

*Attitudes (ATT)*

-Please indicate your level of agreement with the following statements: Being an entrepreneur implies more advantages than disadvantages to me; A career as entrepreneur is attractive for me; If I had the opportunity and resources, I would become an entrepreneur; Being an entrepreneur would be very satisfying for me; Among various options, I would rather become an entrepreneur. (1-7 Likert scale)

*Subjective norms (SUB)*

-If you were to pursue a career as an entrepreneur, how would people in your environment react? Your close family/your friends/your fellow students. (1-7 Likert scale)

*Self-efficacy (SEF)*

-Please indicate your level of competence in performing the following tasks: Identifying new business opportunities; Creating new products and services; Managing innovation within a business; Being a leader and a communicator; Building up a professional network; Commercialising a new idea or development; Successfully managing a business. (1-7 Likert scale)

*Education - University (UNI), Field of study (STU)*

-Please select your university/university of applied science. What is your main field of study?

*Family business background (FAM)*

-Are your parents self-employed or majority owners of a business? No/Yes

## Hypotheses

In our article, we treat the terms 'university entrepreneurial ecosystem' and 'university entrepreneurial environment' as synonymous concepts. In the course of the work, we focus on the perception of students. This means that not the real activity of the universities but rather their perception by the students and their main shaping factors are the focuses of the research. During operationalisation, the concept was understood as the students' perception of a university atmosphere that encourages and supports developing business ideas, starting businesses and engaging in entrepreneurial activities.

There may be several reasons why the entrepreneurial ecosystem of each university differs. Although the curriculum is largely standard in Hungarian universities by field, the exact content of the subjects and the efforts beyond the curriculum (competitions, workshops, business clubs and others) can vary greatly. There is also likely to be a difference in how much money each university spends on such tasks, but Bergman et al. (2018) found no correlation between the amounts spent on entrepreneurial activity and student perceptions of the entrepreneurial climate. At the same time, they found that the size of universities has a negative impact, while its reputation has a positive impact on the entrepreneurial climate. The latter may also be related to the geographical location and the characteristics of the students admitted (admission scores, interest).

The choice of the field of study on the one hand expresses the student's interest, and on the other hand determines the student's curriculum. We know that formal training is a significant explanation for the development of entrepreneurial intentions (Gubik 2013; Szerb & Lukovszki 2013) but it is also decisive in how much and what kind of information reaches students. Participants in entrepreneurship programs are more likely to start their own business (Kolvereid & Moen 1997). The knowledge, experience and confidence – conveyed partly by education and partly by the narrower and wider environment of students – also influence how they perceive their own knowledge and how they see their role in shaping their own future. This, in turn, affects the recognition of opportunities and the efforts made to exploit them. To express this, we use the self-efficacy concept of Bandura (1982), which is defined as people's sense of personal efficacy to produce and to regulate events in their lives. Self-efficacy judgments, whether accurate or faulty, influence peoples' choices and also determine how much effort people will expend and how long they will persist in the face of obstacles or adverse experiences (Bandura 1982). From the viewpoint of this paper entrepreneurial self-efficacy is relevant, which is the "strength of a person's belief that he or she is capable of successfully performing the various roles and tasks of entrepreneurship" (Chen et al. 1998, p. 295).

Our hypotheses are:

H1a: Student perception of the university's entrepreneurial environment varies from university to university.

H1b: The field of study of the students influences the perception of the entrepreneurial environment of the university.

H1c: Self-efficacy influences the perception of the entrepreneurial environment of the university.

In general, research deals with the impact of education on entrepreneurship (Kolvereid & Moen 1997; Chen et al. 1998; Gubik 2013; Szerb & Lukovszki 2013). However, this relationship cannot be construed as a one-way relationship. After all, motivation also strongly influences how much we notice and value the activities that target us. Even in the case of compulsory programs (such as a compulsory university course), we may experience large differences in student performance and activity according to how motivated they are to enter the program. Students who have a positive attitude towards starting a business and who have a high intention to do business are more attentive to the opportunities offered by universities. Thus, they are likely to have a better perception of the entrepreneurial climate and also appreciate the efforts of the university.

H2: The entrepreneurial motivations of the students positively influence the perception of the entrepreneurial environment of the university.

Numerous studies confirm the positive impact of entrepreneurial family background on entrepreneurial intention. Laspita and his colleagues (Laspita et al. 2012) also highlighted that the strength of the effect varies across cultures. Autio and Wennberg (2010) suggest that individuals' community norms and attitudes can have more influence on young people's entrepreneurial behaviour than their own personal attitudes and perceived self-efficacy. Role models emerge as influential factors in individual decision making (Bosma et al. 2012), and thus family entrepreneurial patterns may be dominant in future career plans of students.

H3a: The entrepreneurial family background of students positively influences student perceptions of the university entrepreneurial environment.

H3b: The supportive social background of students positively influences student perceptions of the university entrepreneurial environment.

These factors are closely related, but exploring the relationships between them is not the purpose of this article.

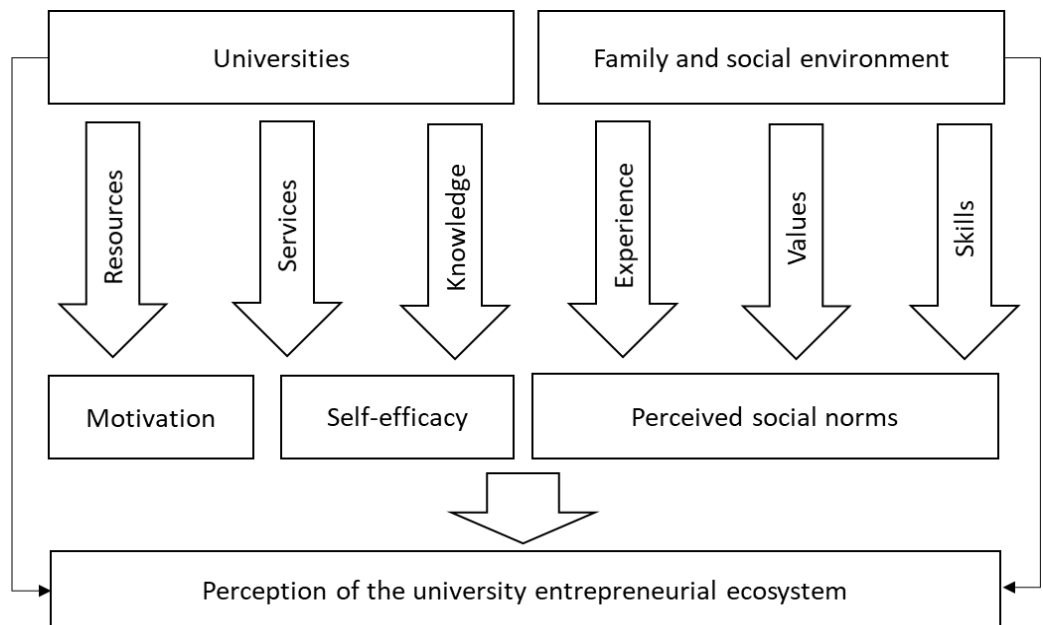

Source: own elaboration

*Figure 1. Perception model of the higher education entrepreneurial ecosystem*

# RESULTS AND DISCUSSION

In the first step, we examined how studies (university, field of study) influence the perception of the university entrepreneurial ecosystem in Hungarian universities. For this, we calculated the average of the responses obtained during the evaluation of the university entrepreneurial environment. This average represents student opinions. First we analysed the differences by university. Only universities with at least 100 respondents (a total of 14 universities) were included in the analysis. We found that there is a weak but significant difference in the perception of the entrepreneurial environment among the students according to the university they attend (Eta Squared = 0.098, p = 0.000).

Based on the assumption that interest is dominant in perception, we assumed that for students in economics/business areas in general value the entrepreneurial environment of universities is higher. The choice of a given profession itself already expresses the interest of the students (and/or the family). It also seems an important argument that economic/business curricula contain a lot of knowledge related to business or entrepreneurial activities. We took this into account when evaluating the ranking per university, by calculating and illustrating the ratio of economics/business students.

Figure 2 shows the university ranking with the differences in the training focus. The left-hand axis shows the perception of the university's entrepreneurial environment (bar chart) and the right-hand axis (line chart) shows the percentage of students in economics/business education among the respondents of the given university. It is clear from the figure that a higher economic/business student ratio is associated with a higher assessment of the environment.

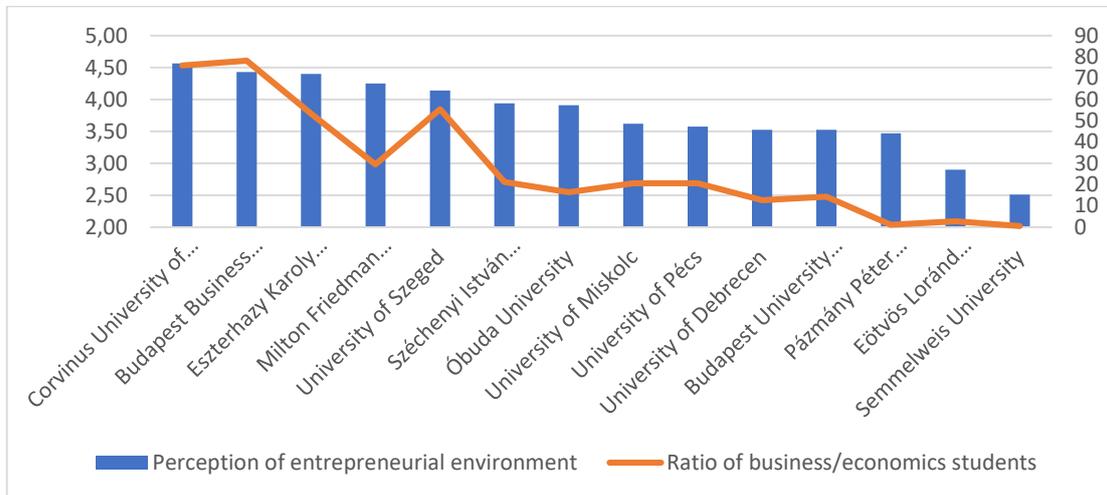

Source: own elaboration, only institutions with above 100 responses. N=9265

*Figure 2. Perception of entrepreneurial ecosystem according to institution*

In the following, we examined the extent to which the field of study affects perception. Figure 3 shows that students in economics/business rate their entrepreneurial environment the highest, followed by social sciences and natural science. The correlation is significant (Eta Squared = 0.102, p = 0.000).

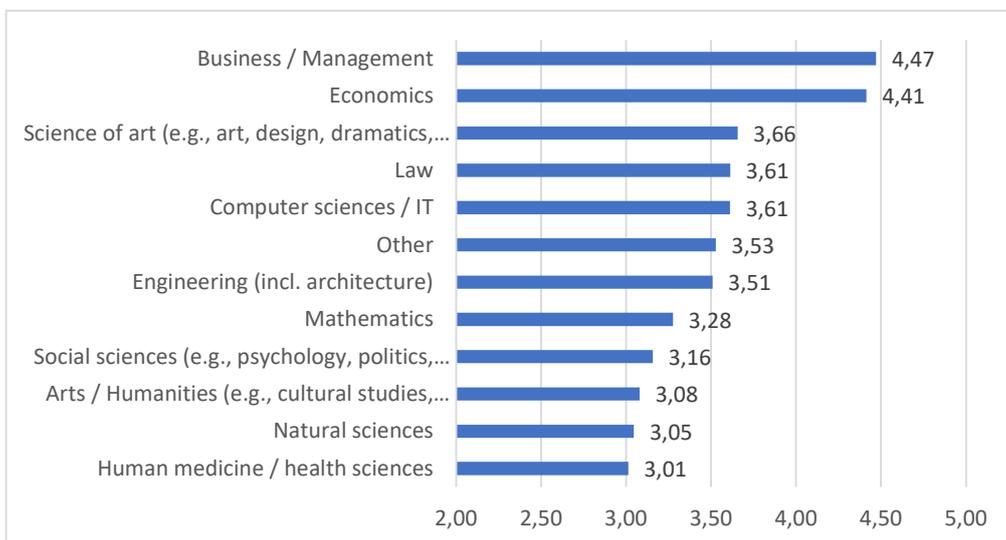

Source: own elaboration, N=9575

*Figure 3. Perception of entrepreneurial ecosystem according to field of study*

After analysing the impact of the field of study on the assessment of the entrepreneurial environment, we repeated the analysis by universities as well. We found that, even when examined on a university-by-university basis, the differences in the assessment of the entrepreneurial environment by field of study remain, but the strength of the relationship lags behind that of the combined calculation.

Another important reason for the differences in the assessment of the university entrepreneurial environment is the entrepreneurial intention of students. Our starting point was that those students with entrepreneurial ideas will pay more attention to the efforts that universities make, such as various programmes, courses or trainings, so they will give a higher value to the entrepreneurial ecosystem.

Table 2 shows that there is a positive and significant correlation between perceptions of the entrepreneurial ecosystem and entrepreneurial intention. The greater the students' entrepreneurial intention, the higher students value the university's entrepreneurial environment (r=0.269, p=0.000). Attitudes toward entrepreneurship also have a positive effect on the evaluation (r=0.236, p=0.000). The more favourable a person's attitude toward entrepreneurship is, the more favourable the evaluation of the university environment is.

Self-efficacy also shows a positive and significant relationship with the entrepreneurial ecosystem

(r=0.371, p=0.000), which can be interpreted as follows: the more an individual feels that he/she has acquired the appropriate skills and knowledge to start an enterprise, the more likely that he/she will give a high rank to the questions.

The last factor is the social norm, which reflects the opinion of the student's environment on the student's entrepreneurial ideas. According to the results (r=0.193, p=0.000) we can conclude that a supportive social environment is also helpful from the viewpoint of the evaluation. Thus, the more positively the individuals' environment reacts to the entrepreneurial plans, the more likely students are to evaluate the university entrepreneurial ecosystem with a high score.

*Table 2*
*Correlation coefficients*

|  | ECO | INT | ATT | SEF | SUB |
|---|---|---|---|---|---|
| ECO | 1 | 0.269** 0.000 | 0.236** 0.000 | 0.371** 0.000 | 0.193** 0.000 |
| INT | 0.269** 0.000 | 1 | 0.845** 0.000 | 0.553** 0.000 | 0.358** 0.000 |
| ATT | 0.236** 0.000 | 0.845** 0.000 | 1 | 0.540** 0.000 | 0.421** 0.000 |
| SEF | 0.371** 0.000 | 0.553** 0.000 | 0.540** .000 | 1 | 0.299** 0.000 |
| SUB | 0.193** 0.000 | 0.358** 0.000 | 0.421** 0.000 | 0.299** 0.000 | 1 |

**Correlation is significant at the 0.01 level (2-tailed). *Correlation is significant at the 0.05 level (2-tailed).
Source: own elaboration

The differences assumed according to the entrepreneurial family background could not be justified. Although the averages of the answers to the ecosystem questions differ slightly (3.57 in the absence of family business background and 3.66 when it exists) the difference was not found to be significant.

## General linear model to evaluate the combined effect

Previously, we checked the pairwise relationships between the variables included in the study. However, we have also seen that the variables used to explain the differences in ecosystem assessment are themselves correlated. In the next step, therefore, we wanted to find out whether the effect of these variables remained significant even if the effects of the other variables were kept under control.

The general linear model (GLM) integrates multivariate linear regression and standard deviation analysis methods (Ketskeméty & Izsó 2005). The dependent variable in our case is the evaluation of the ecosystem, which is derived from the average of the responses measured on the Likert scale from 1 to 7, i.e. it enters the model as a continuous variable. This dependent variable is expressed as a linear function of continuous and discrete independent variables. Here the variables previously described were included in the model, such as university (UNI), field of study (STU), entrepreneurial intention (INT), attitudes (ATT), social norm (SUB) and family entrepreneurial background (FAM). The Type I method takes into account all parameters that cannot be expressed with the other parameters at the same time.

*Table 3*
*Tests of Between-Subjects Effects*

| Source | Type I Sum of Squares | df | Mean Square | F | Sig. |
|---|---|---|---|---|---|
| Corrected Model | 4082.329a | 184 | 22.187 | 12.325 | 0.000 |
| Intercept | 84975.334 | 1 | 84975.334 | 47203.841 | 0.000 |
| SEF | 2207.312 | 1 | 2207.312 | 1226.163 | 0.000 |
| SUB | 85.432 | 1 | 85.432 | 47.458 | 0.000 |

| INT | 61.059 | 1 | 61.059 | 33.919 | 0.000 |
|---|---|---|---|---|---|
| UNI | 959.410 | 23 | 41.713 | 23.172 | 0.000 |
| STU | 346.455 | 11 | 31.496 | 17.496 | 0.000 |
| UNI*STU | 422.661 | 147 | 2.875 | 1.597 | 0.000 |
| Error | 12153.004 | 6751 | 1.800 | | |
| Total | 101210.667 | 6936 | | | |
| Corrected Total | 16235.333 | 6935 | | | |

R Squared = 0.251 (Adjusted R Squared = 0.231)
Source: own elaboration

The fit of the model can be expressed by the value of adjusted R Squared. A value of 0.231 suggests that the variables included in the analysis explain 23.1 per cent of the variability of the data.

In the model, the effect of the intercept and the variables of entrepreneurial intention (INT), self-efficacy (SEF), subjective norm (SUB), university (UNI) and field of study (STU) are significant. Students' assessment of the entrepreneurial ecosystem of universities depends on both the university that students attend, on the major they have, and on a combination of these two variables. The difference according to the universities is significant when the training area is under control. Although in the framework of this article we did not examine the reason for the different evaluation of universities, it can be assumed that it is to be found in the different efforts of universities, geographical reasons, and the specifics of the student body. The field of study can have the greatest impact through the curriculum, as the economic and entrepreneurial subjects that students may encounter are very different from field to field. The more an individual feels that she possesses the skills and knowledge needed to start a business, the higher she rates the entrepreneurial ecosystem of the university. This is self-efficacy, which complements the two objective variables related to studies with the subjective assessment of knowledge by students. Based on the results, Hypotheses H1a, b and c can be accepted.

The significant effect of entrepreneurial intention on the ecosystem assessment suggests that students who plan to start a business in the future will value the entrepreneurial efforts of universities more highly. The attitude variable could not substantially increase the explanatory power of the model, presumably due to its close correlation with entrepreneurial intention. The results led to the acceptance of Hypothesis H2. It is also decisive whether students come from an entrepreneur-friendly environment (feedback from their narrower and wider environment is positive toward entrepreneurship), this is the subjective norm. A student with such an environment is likely to value the entrepreneurial knowledge and experience gained at the university even more.

The family entrepreneurial background did not prove to be significant in the model, so it is not included in the final version presented here. However, its combined effect with the university and the field of education variables is significant (although their partial effect is very weak). This may suggest that the impact of family entrepreneurship background may be reflected in the choice of university and field of education. This assumption requires further investigation. We were able to confirm Hypothesis H3b using the calculations, but we could not support Hypothesis H3a.

# CONCLUSION

The main focus area of entrepreneurship education studies is the impact of entrepreneurship courses/training on the entrepreneurial intentions of students. Most studies find a positive relationship between the two, which suggests that universities can boost entrepreneurship through standard courses and extracurricular academic programmes. But the results are not unanimous. Some studies find no significant relationship between the two phenomena, and there are even some that suggest a negative relationship. Our findings point to a positive relationship: the field of study is positively associated with the students' evaluation of the entrepreneurial ecosystem, while there is also a positive correlation between the entrepreneurial ecosystem evaluation and the entrepreneurial intentions. Students enrolled in business/economics study programmes in Hungary are more likely to have entrepreneurship courses, and so their evaluation on the entrepreneurial ecosystem is higher; a higher evaluation of the ecosystem is also positively correlated with higher entrepreneurial intentions.

The perspective of our study, however, is different from that usually taken in the literature. We focus on student perception, namely how students evaluate the entrepreneurial ecosystem around the university, and we suggest that the perception is influenced by a number of factors, including the entrepreneurial intentions of students. In this sense our perspective is similar to the one taken by Bergmann et al. (2018). In our opinion it is not enough for universities to focus only on what programmes and services they come up with in order to develop the students' entrepreneurial skills. Our analyses has shown that a number of individual student factors, such as their prior knowledge of entrepreneurship, their family background, how their environment relates to entrepreneurship, and their area of interest largely determine whether they perceive opportunities at all. Students who perceive opportunities are presumably more likely to take advantage of them by participating in non-compulsory programmes, as well as being more engaged in required entrepreneurship courses.

The general linear model we set up (using a database of almost 10,000 Hungarian university student answers obtained in 2018) found that the 1) entrepreneurial intentions of students, 2) their self-efficacy (the rate of confidence that the student possesses the skills that are needed to start a business), 3) the social norms perceived by the students, 4) the university they attend, and 5) the field of study all have a small but statistically significant impact on how students perceive the entrepreneurial ecosystem around the university. These five factors explain 23.1% of the variability of the data.

It seems that along with their efforts to improve entrepreneurial skills, universities should pay much more attention to shaping attitudes and arousing students' interest in entrepreneurship. Many higher education institutions are making serious efforts to develop their entrepreneurial character by launching various programmes, providing services and, more generally, by creating an entrepreneurial environment. Very often the impact of these programs is below expectations. The goal is not only to further increase the number of courses or provision of services, but to awaken student demand for them.

*Acknowledgement*

*The research was supported by project number EFOP-3.6.2-16-2017-00007, entitled "Aspects on the development of intelligent, sustainable and inclusive society: social, technological and innovation networks in employment and digital economy". The project is implemented with the support of the European Union, co-financed by the European Social Fund and the state budget of Hungary.*